\documentclass[aip,apl,reprint]{revtex4-1}

\pdfoutput=1
\usepackage{graphicx}
\usepackage{amsmath}

\begin{document}


\title{A micro-structured ion-implanted magnonic crystal} 



\author{Bj\"orn~Obry}
\author{Philipp~Pirro}
\affiliation{Fachbereich Physik and Forschungszentrum OPTIMAS, Technische Universit\"at Kaiserslautern, D-67663 Kaiserslautern, Germany}

\author{Thomas~Br\"acher}
\affiliation{Fachbereich Physik and Forschungszentrum OPTIMAS, Technische Universit\"at Kaiserslautern, D-67663 Kaiserslautern, Germany}
\affiliation{Graduate School Materials Science in Mainz, D-67663 Kaiserslautern, Germany}

\author{Andrii~V.~Chumak}
\affiliation{Fachbereich Physik and Forschungszentrum OPTIMAS, Technische Universit\"at Kaiserslautern, D-67663 Kaiserslautern, Germany}

\author{Julia~Osten}
\affiliation{Institut f\"ur Ionenstrahlphysik und Materialforschung, Helmholtz-Zentrum Dresden-Rossendorf, D-01328 Dresden, Germany, and Technische Universit\"{a}t Dresden, D-01062 Dresden, Germany}

\author{Florin~Ciubotaru}
\author{Alexander~A.~Serga}
\affiliation{Fachbereich Physik and Forschungszentrum OPTIMAS, Technische Universit\"at Kaiserslautern, D-67663 Kaiserslautern, Germany}

\author{J\"urgen~Fassbender}
\affiliation{Institut f\"ur Ionenstrahlphysik und Materialforschung, Helmholtz-Zentrum Dresden-Rossendorf, D-01328 Dresden, Germany, and Technische Universit\"{a}t Dresden, D-01062 Dresden, Germany}

\author{Burkard~Hillebrands}
\affiliation{Fachbereich Physik and Forschungszentrum OPTIMAS, Technische Universit\"at Kaiserslautern, D-67663 Kaiserslautern, Germany}


\date{\today}

\begin{abstract}
We investigate spin-wave propagation in a microstructured magnonic-crystal waveguide fabricated by localized ion implantation. The irradiation caused a periodic variation in the saturation magnetization along the waveguide. As a consequence, the spin-wave transmission spectrum exhibits a set of frequency bands, where spin-wave propagation is suppressed. A weak modification of the saturation magnetization by $7\%$ is sufficient to decrease the spin-wave transmission in the band gaps by a factor of 10. These results evidence the applicability of localized ion implantation for the fabrication of efficient micron- and nano-sized magnonic crystals for magnon spintronic applications.
\end{abstract}


\maketitle 

Magnonic crystals belong to the class of metamaterials and are artificially patterned magnetic media. This manufactured periodic pattern results in the formation of forbidden frequency bands for spin-wave excitations of the material.\cite{Sykes1976, Nikitov2001} The feasibility to store or process information by spin waves using a magnonic crystal\cite{Chumak2010, Chumak2012} makes it a promising tool for magnon spintronic applications.\cite{Kruglyak2010} The future use in a chip-based spin-logic architecture, however, requires high-quality microscopic magnonic crystals that are easy to produce. Presently, they are realized in the form of microscopic waveguides with variable width,\cite{Chumak2009} arrays of coupled magnetic strips,\cite{Gubbiotti2007, Topp2010} antidot lattices\cite{Lenk2012} and bicomponent magnonic crystals,\cite{Wang2009, Tacchi2012} all of which have proven as reliable methods. But there appear drawbacks like the formation of higher modes at the edge steps of the modulated sample topography\cite{Lee2009, Chumak2009, Ciubotaru2012} and the complicated fabrication procedures, respectively.

These problems can be overcome by patterning the magnetic properties of a waveguide without major changes of the sample topography by localized ion implantation.\cite{Volluet1981, Carter1982, Kaminsky2001, Barsukov2011} Irradiating a Ni$_\text{81}$Fe$_\text{19}$ (permalloy) film with ions leads to a reduction of its saturation magnetization.\cite{Folks2003, Fassbender2006, Fassbender2006b} Thus, localized ion implantation can be utilized to manipulate the spin-wave propagation by a pure magnetic patterning of a permalloy film.\cite{Obry2013}

In this letter we present a one-dimensional micro-structured magnonic crystal waveguide with a periodic modulation of the saturation magnetization obtained by localized ion implantation. While the magnetic patterning method has nearly no impact on the waveguide topography, the change in the spin-wave propagation in the waveguide is drastic. A set of pronounced frequency band gaps exists in the spin-wave transmission spectrum, corresponding to a rejection of spin waves by the magnonic crystal. Furthermore, no scattering between different transversal waveguide modes occurs at the boundaries of the periodic pattern, as was also previously reported for other microscopically structured magnonic waveguides.\cite{Chumak2009, Lee2009, Ciubotaru2012}

\begin{figure}
	\includegraphics[viewport = 67 510 500 770, clip, scale=1.0, width=1.0\columnwidth]{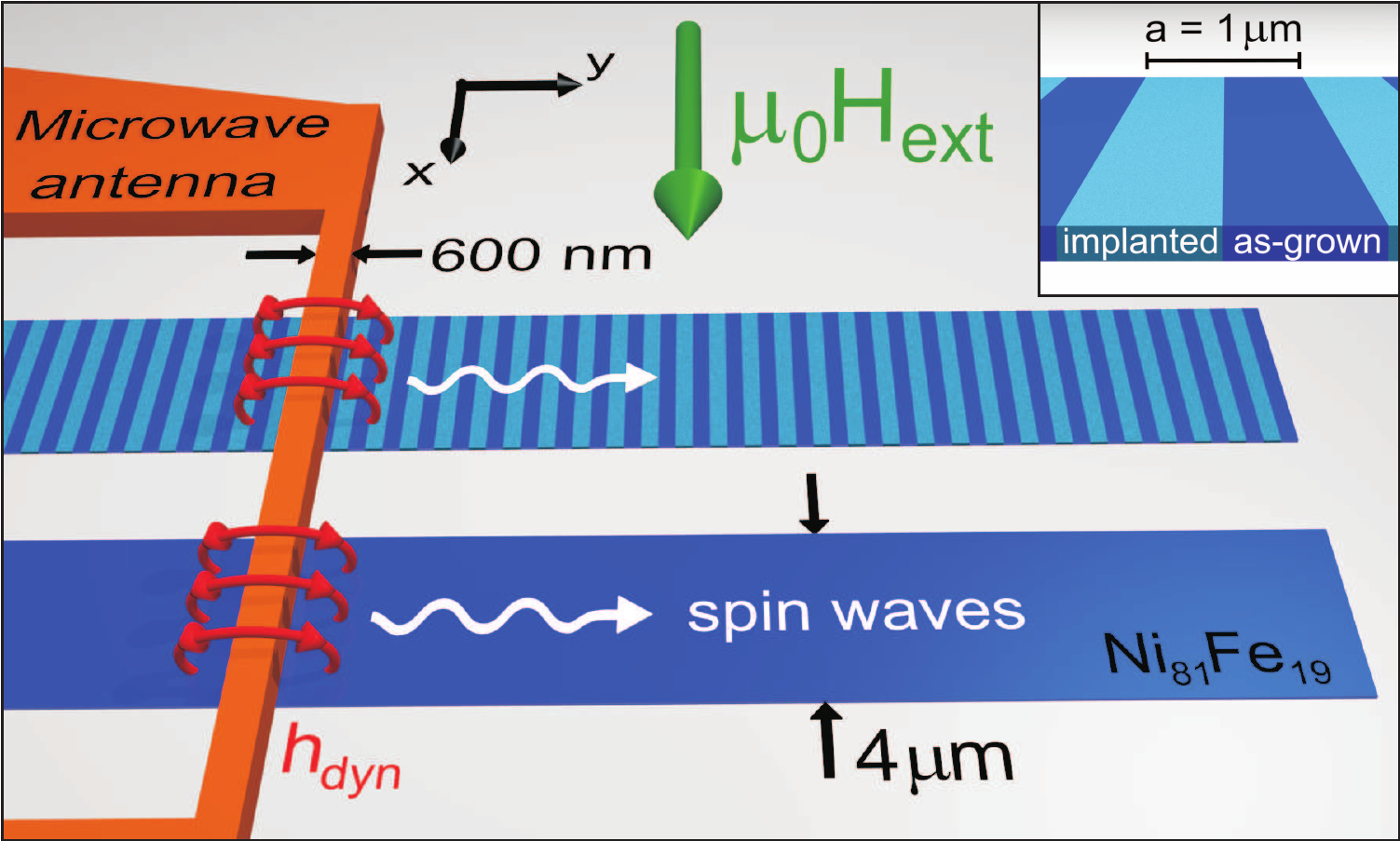}
	\caption{\label{Fig1} (Color online) Experimental sample setup. Spin waves are excited in two $4\,\mu$m wide Ni$_\text{81}$Fe$_\text{19}$ waveguides by the dynamic Oersted field $h_\text{dyn}$ of a $400$\,nm thick and $600$\,nm wide Cu antenna. Along one of the waveguides the saturation magnetization has been modulated periodically by localized Cr$^\text{+}$ implantation with a lattice constant $a = 1 \mu$m (see inset). The other waveguide acts as a reference. An external in-plane magnetic field $H_\text{ext}$ is applied perpendicularly to the spin-wave propagation direction in order to saturate the magnetization.}
\end{figure}

The experiments were performed on two Ni$_\text{81}$Fe$_\text{19}$ waveguides (Fig.\,\ref{Fig1}) with a width of $w = 4\,\mu$m and a nominal thickness of $t = 20$\,nm, which were fabricated by means of conventional lift-off techniques and molecular beam evaporation. While one of the waveguides served as a reference waveguide, a periodic modulation in the saturation magnetization $M_\text{S}$ was imposed on the second waveguide by localized ion implantation. For this, a polymethyl methacrylate (PMMA) mask was patterned on top of the sample by e-beam lithography containing a set of $4\,\mu$m wide and $500\,$nm long windows with a periodicity of $1\,\mu$m (inset of Fig.\,\ref{Fig1}). The subsequent irradiation of the sample by a broad beam of Cr$^\text{+}$ ions with a kinetic energy of $30\,$keV and a fluence of $4.4\cdot 10^{15}$\,ions/cm$^2$ resulted in a decrease of $M_\text{S}$ in the unshielded regions to $(93\pm 6)\%$ of its original value, as determined from polar magneto-optic Kerr effect measurements\cite{Marko2010} on equally treated reference samples. Finally, a $200$\,nm thick and $600$\,nm wide Cu antenna was produced on top of the waveguides (Fig.\,\ref{Fig1}). Applying a microwave current to the antenna, spin waves were excited in the Ni$_\text{81}$Fe$_\text{19}$ waveguides by the dynamic Oersted field $h_\text{dyn}$ of the antenna, and they propagated perpendicularly to an external in-plane magnetic field $\mu_\text{0}H_\text{ext} = 30$\,mT (Damon-Eshbach geometry). The spin-wave intensity was detected using Brillouin light scattering (BLS) microscopy, which provides a high dynamic spectral range and a spatial resolution of $250$\,nm.\cite{Demidov2004}

\begin{figure}
	\includegraphics[viewport = 98 439 452 727, clip, scale=1.0, width=0.8\columnwidth]{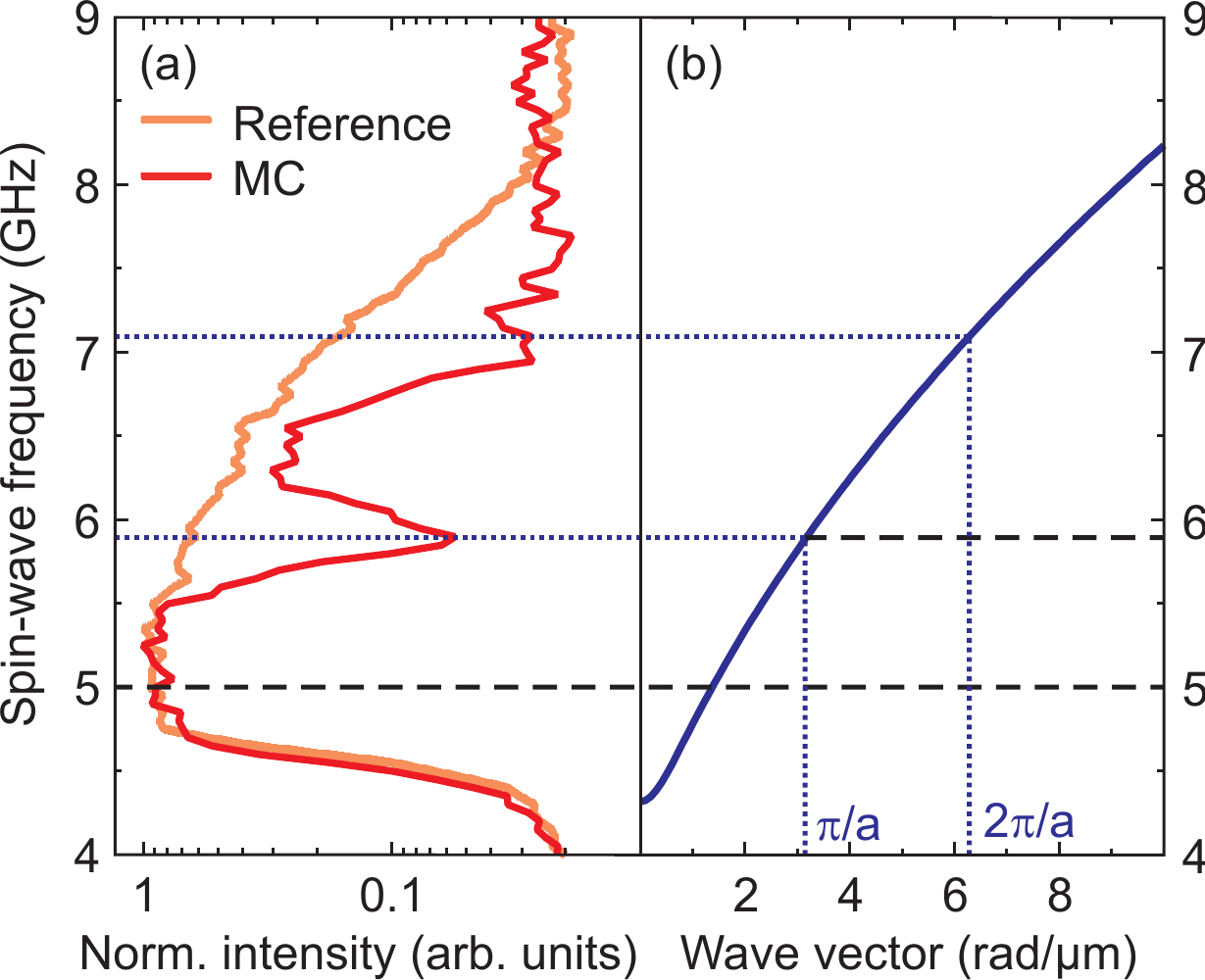}
	\caption{\label{Fig2} (Color online) (a) Normalized spin-wave transmission spectra of the magnonic crystal waveguide (dark line) and the reference waveguide (pale line). The spectra were recorded for an external magnetic field of $\mu_\text{0}H_\text{ext} = 30$\,mT. (b) Calculated spin-wave dispersion relation for the same field and a waveguide width of $4\,\mu$m. The first two band gap positions are indicated by the dotted lines. Dashed lines mark spin-wave frequencies studied in Fig.\,\ref{Fig3}.}
\end{figure}

The magnonic crystal character of the modulated waveguide becomes apparent in the measurement of the spin-wave transmission spectra (Fig.\,\ref{Fig2}(a)). Here, the detected spin-wave intensity measured at a propagation distance of a few micrometers from the antenna has been plotted as a function of the microwave excitation frequency. For a better comparison, the spectra were normalized in intensity. With respect to the reference waveguide, the transmission spectrum of the magnonic crystal waveguide reveals two pronounced band gaps for spin-wave frequencies at $5.9$\,GHz and at about $7.0$\,GHz, where the spin-wave intensity drops by a factor of $10$. These frequency regions constitute the rejection bands of the magnonic crystal. 

Figure\,\ref{Fig2}(b) displays the Damon-Eshbach type spin-wave dispersion relation, which was calculated according to Ref.\,\onlinecite{Kalinikos1986} for a magnetic field of $\mu_\text{0}H_\text{int} = 27.3$\,mT and a saturation magnetization of $M_\text{S} = 675$\,kA/m.\cite{parameters1} The latter value was determined from polar magneto-optic Kerr effect measurements on non-irradiated Ni$_\text{81}$Fe$_\text{19}$ reference films, while the magnetic field value takes into account the demagnetizing fields originating at the waveguide edges.\cite{Demokritov2009} The dispersion relation reflects the characteristic features of the reference transmission spectrum. While the strong onset of spin-wave transmission at lower frequencies corresponds to the ferromagnetic resonance (FMR) frequency $f_\text{FMR} = 4.3$\,GHz, the gradual decrease in the detected spin-wave intensity towards higher frequencies can be understood by taking into account the finite size of the microwave antenna.\cite{Demidov2009} For the given antenna width of $600$\,nm an efficient excitation of spin waves is restricted to spin waves up to a maximum wave vector of $k_\text{max} \approx 10\,\text{rad}/\mu\text{m}$. According to Fig.\,\ref{Fig2}(b), this results in an upper limit for the spin-wave frequency of $f_\text{max} = 8.2$\,GHz, which perfectly agrees with the observed values in the transmission spectrum of the reference waveguide (Fig.\,\ref{Fig2}(a)).

Furthermore, for the transmission spectrum of the magnonic crystal waveguide, the positions of the frequency band gaps coincide with the first two resonance conditions for Bragg scattering from the periodic magnonic crystal pattern. The latter are indicated by the dotted lines in Fig.\,\ref{Fig2} and were calculated using the Bragg condition $k = \frac{n\pi}{a}$, which connects the spin-wave wave vector $k$ with the lattice constant $a = 1\,\mu$m of the artificial lattice, where n is an integer number. It has to be noted that the waveguide thickness $t$ was used as a fitting parameter in the calculation of the spin-wave dispersion relation (Fig.\,\ref{Fig2}(b)). While the thickness has basically no impact on the position of the FMR frequency, it strongly influences the frequency values for Bragg reflection. The resulting thickness value of $t = 19$\,nm reproduces the nominal thickness quite well. Since no capping layer was used on top of the Ni$_\text{81}$Fe$_\text{19}$, the oxidation of the waveguide surface and thus a reduction of the effective thickness is most likely.

\begin{figure}
	\includegraphics[viewport = 27 222 393 720, clip, scale=1.0, width=1.0\columnwidth]{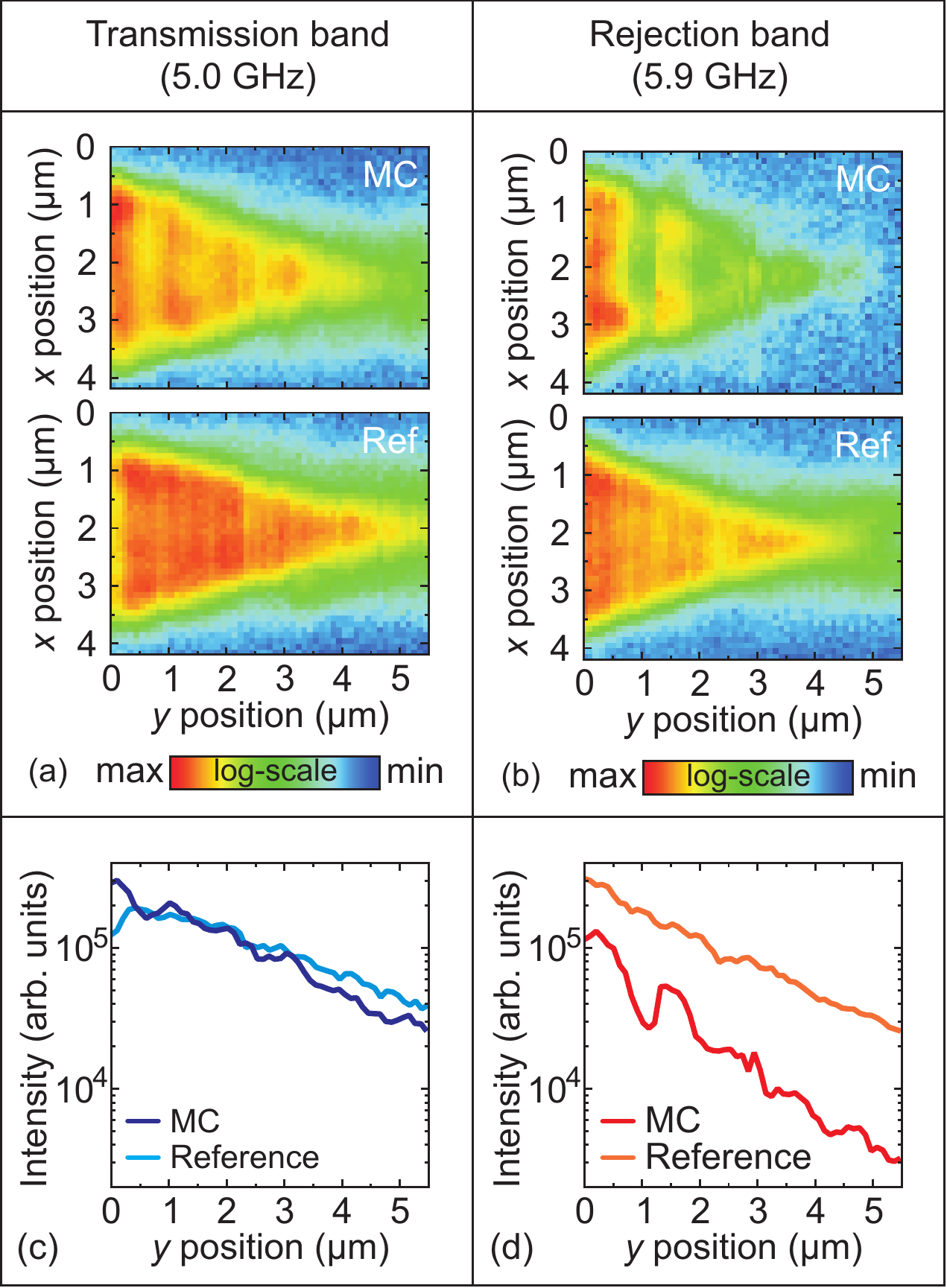}
	\caption{\label{Fig3} (Color online) Investigation of the spatial spin-wave intensity distribution with a frequency in the transmission band ((a) and (c)) and in the rejection band ((b) and (d)). The lateral spin-wave intensity distribution was mapped by Brillouin light scattering microscopy for the magnonic crystal waveguide (``MC'', upper graphs in (a) and (b)) and the reference waveguide (``Ref'', lower graphs in (a) and (b)). For a comparison of the spatial decay along the waveguides the integral intensity distribution along the waveguide is plotted ((c) and (d)). The microwave antenna is located at $y = 0\,\mu$m.}
\end{figure}

A more detailed analysis of the spin-wave propagation in the magnonic crystal waveguide was achieved by space-resolved BLS microscopy measurements (Fig.\,\ref{Fig3}). Here, spatial maps of the spin-wave intensity distribution within the waveguides were taken for an excitation frequency in the transmission and rejection band, respectively. The investigated frequencies are indicated by the dashed lines in Fig.\,\ref{Fig2}. Spin waves were excited by the antenna at $x = 0\,\mu$m and detected locally by scanning the BLS laser focus across the waveguide. The detected spin-wave intensity is color-coded with red (blue) color representing high (low) intensity. In both cases the propagation in the magnonic crystal waveguide was compared to the propagation in the conventional reference waveguide. 

For the excitation of spin waves in the transmission band with a frequency of $f_\text{exc} = 5.0$\,GHz a weak difference in both intensity maps was observed (Fig.\,\ref{Fig3}(a)). While in the reference waveguide the spin waves decay homogeneously with increasing distance from the antenna, a faster decay as well as a small overlaid oscillation in the intensity occurs in the magnonic crystal waveguide. Apparently, this is due to the implementation of the irradiated regions with a reduced $M_\text{S}$ and an increased Gilbert damping\cite{Fassbender2006} allowing for spin-wave reflection at the transitions between the irradiated and non-irradiated areas. A more quantitative analysis of the spin-wave decay along the waveguide is given in Fig.\,\ref{Fig3}(c). Here, the total transmitted spin-wave intensity is shown as a function of the distance from the antenna, obtained by integrating the detected spin-wave signal over the $x$ position in Fig.\,\ref{Fig3}(a). In the case of the reference waveguide the slope of this curve yields information about the spin-wave damping. However, for the magnonic crystal waveguide this is not trivial, since multiple reflections from the irradiated regions as well as the periodically modified saturation magnetization and damping have to be taken into account. In order to simplify the analysis, an effective decay length $d$ was used to quantify the transmission through the waveguide, i.e., to account for both damping and reflection losses. Fitting an exponential decay of the spin-wave intensity $I$ according to $I(y) = I(y=0)\cdot exp(-y/d)$ to the experimental data yielded values of $d = 3.2\,\mu$m for the reference waveguide and $d = 2.6\,\mu$m for the magnonic crystal waveguide (Fig.\,\ref{Fig3}(c)). A small increase in $d$ by a factor of $1.2$ due to the irradiation was determined.

In contrast, the excitation of spin waves with a frequency of $f_\text{exc} = 5.9$\,GHz, i.e., inside the band gap, has a notable impact on their propagation in the magnonic crystal waveguide. The intensity maps in Fig.\,\ref{Fig3}(b) reveal a pronounced overlaid oscillation in the magnonic crystal waveguide being a sign of a standing spin-wave pattern due to the Bragg reflections from the artificial lattice. In addition, the decay of the transmitted intensity (Fig.\,\ref{Fig3}(d)) in the magnonic crystal waveguide ($d = 1.2\,\mu$m) was increased with respect to the reference waveguide ($d = 2.1\,\mu$m) by a factor of $1.8$. Beside the increased effective damping, the offset in intensity between both curves in Fig.\,\ref{Fig3}(d) suggests that the excitation of spin waves in the magnonic crystal waveguide within the band gap has a lower efficiency. These observations give rise to the assumption that the suppression of spin waves by a factor of 10 in the rejection band is caused by two different effects: Partly, the effective spin-wave damping is increased due to Bragg scattering from the artificial crystal, but additionally the location of the antenna inside the magnonic crystal leads to a decreased excitation efficiency inside the rejection band. 

In conclusion, localized ion implantation is a convenient method to fabricate high-performance magnonic crystal waveguides on the microscopic scale. A relatively weak, periodic change in the waveguide's saturation magnetization of $7\%$ is sufficient for the formation of pronounced band gaps, where spin-wave propagation is efficiently suppressed. Since the waveguide consisted of one single, topographically unchanged material, spin waves with frequencies in the transmission bands were nearly unaffected. Hence, this class of magnonic crystal waveguides bears a high potential for future application in magnon spintronic devices.

The authors would like to thank the Nano Structuring Center (NSC) of the TU Kaiserslautern for support with the sample preparation and E. Papaioannou for the evaporation of the Ni$_\text{81}$Fe$_\text{19}$ films. Financial support by the DFG (SE-1771/1-2) is gratefully acknowledged.



%
%

%




\end{document}